325.tex

J.Inverse and Ill-Posed problems 2,N3,(1994),269-275.

\input amstex
\documentstyle{amsppt}
 
\pagewidth{6.4 truein}
\pageheight{8.6 truein}
\hfuzz=10pt
 
 
\TagsOnRight
 
\topmatter
\title Stability of the Solution to Inverse Obstacle Scattering Problem
\endtitle
\rightheadtext{Stability of the Solution}
\author A.G. Ramm \endauthor
\affil  Department of Mathematics, Kansas State University,
        Manhattan, KS  66506-2602, USA \endaffil
 
\abstract {It is proved that if the scattering amplitudes for two
obstacles (from a large class of obstacles) differ a little, then the
obstacles differ a little, and the rate of convergence is given.  An
analytical formula for calculating the characteristic function of the
obstacle is obtained, given the scattering amplitude at a fixed
frequency.} \endabstract
 
\endtopmatter
 
\vglue .2in
 
\document
 
\subhead Introduction \endsubhead
 
Let  $D \subset {\Bbb R}^3$  be a bounded domain with a smooth boundary
$\Gamma$,
$$
(\nabla^2 + k^2)u = 0 \quad\text{in}\quad D^\prime := {\Bbb R}^3
\setminus D, \quad k = \text{const} > 0; \quad u = 0
\quad\text{on}\quad \Gamma \tag 1
$$
$$
u = \exp (ik \alpha \cdot x) + A(\alpha^\prime, \alpha, k) r^{-1} \exp
(ikr) + o(r^{-1}), \quad r := |x| \to \infty, \quad \alpha^\prime :=
xr^{-1}.  \tag 2
$$
Here  $\alpha \in S^2$  is a given unit vector, $S^2$  is the unit
sphere in  ${\Bbb R}^3$, the function  $A(\alpha^\prime, \alpha, k)$  is
called the scattering amplitude (the radiation pattern).  It is well
known [1] that problem (1)-(2) has a unique solution, the scattering
solution, so that the map  $\Gamma \to A(\alpha^\prime, \alpha, k)$  is
well defined.  We consider the inverse obstacle scattering problem
(IOSP): {\it given  $A(\alpha^\prime, \alpha) := A(\alpha^\prime,
\alpha, k = 1)$  for all  $\alpha^\prime, \alpha \in S^2$  and a fixed
$k$  (for example, take  $k = 1$  without loss of generality), find
$\Gamma$.}
 
Let us assume that  $\Gamma \subset \gamma_\lambda$, where
$\gamma_\lambda$  is the set of star-shaped (with respect to a common
point  $O$) surfaces, which are located in the annulus  $0 < a_0 \leq
|x| \leq a_1$, and whose equations  $x_3 = \phi (x_1,x_2)$  in the local
coordinates (in which  $x_3$  is directed along the normal to  $\Gamma$
at a point  $s \in \Gamma$), have the property
$$
\| \phi \|_{C^{2,\lambda}} \leq c_0, \tag 3
$$
$C^{2,\lambda}$  is the space of twice differentiable functions, whose
second derivatives satisfy the H\"older condition of order  $0 < \lambda
\leq 1$, $\lambda$  and  $c_0$  are independent of  $\phi$  and
$\Gamma$.
 
Uniqueness of the solution to IOSP with fixed frequency data is first
proved in [1, p.\ 85].  We are interested here in the stability problem:
suppose that  $\Gamma_j \in \gamma _\lambda$
  generate  $A_j (\alpha^\prime, \alpha)$, $j = 1,2$, and
$$
\max_{\alpha^\prime, \alpha \in S^2} |A_1 (\alpha^\prime, \alpha) - A_2
(\alpha^\prime, \alpha)| < \delta.  \tag 4
$$
What can one say about the Hausdorff distance between  $D_1$  and
$D_2$: $\rho := \sup_{x \in \Gamma_1}\inf_{ y \in \Gamma_2}|x-y|$.
Let  $\tilde D_1$  denote a connected component of  $D_1 \setminus D_2$,
$D_{12} := D_1 \cup D_2$, $\Gamma_{12} := \partial D_{12}$,
$D_{12}^\prime :=\Bbb R^3\setminus D_{12},$
 $\tilde\Gamma_1 := \partial \tilde D_1 := \Gamma_1^\prime \cup \tilde
\Gamma_2$, $\tilde \Gamma_2 \subset \Gamma_2 := \partial D_2$,
$\Gamma_1^\prime \subset \Gamma_1 := \partial D_1$.  Let us assume,
without loss of generality, that  $\rho = |x_0 - y_0|$, $x_0 \in
\Gamma_1^\prime$, $y_0 \in \tilde \Gamma_2$.
Can one obtain a formula for calculating  $\Gamma$, given
$A(\alpha^\prime, \alpha)$  for all  $\alpha^\prime, \alpha \in S^2$, $k
= 1$  is fixed?  No such formula is known for IOSP.  For inverse
potential scattering problem with fixed-energy data such a formula and
stability estimates are obtained in [2], [3]. These results are based
on the works [7],[8], [10]-[17], [19]-[21].
 
In section II we prove that  $\rho \leq c_1 \left( \frac {\ln | \ln
\delta |}{|\ln \delta|}\right)^{c_2}$  as  $\delta \to 0$.  We also
prove some inversion formula, but it is an open problem to make an
algorithm out of this formula.  In Remark 3, we comment on some recent
papers [4-6] in which attempts are made to study the stability problem
and point out a number of errors in these papers. Our result,
formulated as Theorem 1 in section II, is stronger than the results
announced in Theorem 1 in [4], Theorem 1 in [5] and Theorem 2.10 in [6].
 
\vskip .2in
 
\subhead II. Stability Result and a Reconstruction Formula \endsubhead
 
\proclaim{Theorem 1} Under the assumptions of section I, one has
$\rho (\delta) \leq c_1 \left( \frac {\ln | \ln \delta |}{|\ln
\delta|}\right)^{c_2},$ where $c_1$ and $c_2$ are positive constants
independent of $\delta$.
\endproclaim
 
\vskip .2in
 
\proclaim{Proposition 1} There exists a function  $\nu_\epsilon (\alpha,
\theta) \in L^2 (S^2)$  such that
$$
-4\pi \lim_{\epsilon \to 0} \int_{S^2} A(\theta^\prime, \alpha)
\nu_\epsilon (\alpha, \theta) d\alpha = -\frac {\lambda^2}2 \tilde
\chi_D (\lambda).  \tag 5
$$
Here  $\lambda \in {\Bbb R}^3$  is an arbitrary fixed vector, $\chi_D
(x) := \cases 1, & x \in D \\ 0, & x \not\in D \endcases$,
$\tilde \chi_D (\lambda) := \int_{{\Bbb R}^3} \exp
(-i\lambda \cdot x) \chi_D (x)dx$, $\theta, \theta^\prime \in M := \{
\theta : \theta \in {\Bbb C}^3, \theta \cdot \theta = 1\}$,
$\theta^\prime - \theta = \lambda$, and  $A(\theta^\prime, \alpha)$
is defined by the absolutely convergent series
$$
A(\theta^\prime, \alpha) = \sum_{\ell = 0}^\infty A_\ell (\alpha) Y_\ell
(\theta^\prime), \quad \theta^\prime \in M, \quad
A_\ell (\alpha) := \int_{S^2} A(\alpha^\prime, \alpha) \overline{Y_\ell
(\alpha^\prime)} d\alpha^\prime, \tag 6
$$
where  $Y_\ell (\alpha)$  are the orthonormal in  $L^2 (S^2)$  spherical
harmonics, $Y_\ell (\theta^\prime)$  is the natural analytic
continuation of  $Y_\ell (\alpha^\prime)$  from  $S^2$  to  $M$, and the
series (6) converges absolutely and uniformly on compact subsets of
$S^2 \times M$.
\endproclaim
 
\vskip .2in
 
\demo{Remark 1} The stability result given in Theorem 1 is similar to
the one in [3], p.\ 9, formula (2.42), for inverse potential scattering.
\enddemo
 
\vskip .1in
 
\demo{Remark 2} Proposition 1 claims the existence of the inversion
formula (5).  An open problem is to construct the function
$\nu_\epsilon (\alpha, \theta)$  algorithmically, given the data  $A
(\alpha^\prime, \alpha) \quad \forall \alpha^\prime, \alpha \in S^2$.
\enddemo
 
\vskip .1in
 
\demo{Proof of Theorem 1} First, we prove that  $\rho (\delta) \to 0$
as  $\delta \to 0$.  Then, we prove that  $|u_2| \leq c\rho$  in
$\tilde D_1$.  Next, we prove that  $|u_2 (x)| \leq
c\epsilon^{\rho^{c^\prime}} \
(\ast)$  if  $\text{dist} (x, \Gamma_1^\prime) = O(\rho)$, where  $|\ln
\epsilon | = cN(\delta)$, $N(\delta) := |\ln \delta |/\ln | \ln \delta
|$.  From  $(\ast)$  Theorem 1       follows.  By  $c$, $c^\prime$,
$\tilde c$,
$c_j$  various positive constants, independent of  $\delta$  and on
$\Gamma \in \gamma_\lambda$, are denoted.
 
\vskip .2in
 
\noindent {\bf Step 1.} {\it Proof of the relation  $\rho (\delta) \to
0$  as  $\delta \to 0$.}  Assume the contrary:
$$
\rho_n := \rho (\delta_n) \geq c > 0 \quad\text{for some sequence}\quad
\delta_n \to 0.  \tag 7
$$
Let  $\Gamma_{jn}$, $j = 1,2$, be the corresponding sequences of the
boundaries, $\Gamma_{jn} \in \gamma_\lambda$.  Due to assumption (3),
one can select a convergent in  $C^{2,\mu}$, $0 < \mu < \lambda$,
subsequence,
which we denote  $\Gamma_{jn}$  again.  Thus  $\Gamma_{jn} \to
\Gamma_j$  as  $n \to \infty$.  From (7) it follows that  $(\dagger) \
\rho
(D_1,D_2) \geq c > 0$, where  $D_j$  is the obstacle with the boundary
$\Gamma_j$.  By the known continuity of the map  $\Gamma_j \to A_j$,
$\Gamma_j \in \gamma_\mu$, it follows that  $A_1 (\alpha^\prime,
\alpha)
- A_2 (\alpha^\prime, \alpha) = 0$.
 
By the uniqueness theorem [1, p.\ 85] it follows that  $\Gamma_1 =
\Gamma_2$.  Thus, $\rho (D_1, D_2) = 0$  which is a contradiction to
$(\dagger)$.  This contradiction proves that  $\rho (\delta) \to 0$  as
$\delta \to 0$.
 
\vskip .2in
 
\noindent {\bf Step 2.} {\it Proof of the estimate  $|u_2 (x)| \leq
c\rho$  for  $x \in \tilde D_1$.}  It is known that  $\| u_2\|_{C^2
(D_2^\prime)} \leq c$, where  $u_2 = u_2 (x, \alpha)$  is the scattering
solution corresponding to the obstacle  $D_2$.  Since  $u_2 = 0$  on
$\tilde \Gamma_2$, one has  $|u_2 (x)| \leq (\max_{x \in \tilde D_1}
|\nabla u_2|) \rho \leq c\rho$.
 
\vskip .2in
 
\noindent {\bf Step 3.} {\it Proof of the estimate  $|v(x)| \leq
c\epsilon^{d^{c^\prime}}$, where
$v:= u_2- u_1$ and $d := \text{dist} (x, \Gamma_1^\prime)$.}

From [3, p.\ 26, formulas (4.12), (4.17), (2.28)], one has
$$
|v(x)| \leq \epsilon := c \exp \{ -\gamma N(\delta)\}, \quad |x| > a_2,
\quad N(\delta) := \frac {|\ln \delta|}{\ln | \ln \delta|}, \quad
\gamma := \ln \frac {a_2}{a_1} > 0,  \tag 8
$$
$a_2 > a_1 $  is an arbitrary fixed number, $a_2 \leq |x| \leq
a_2 + 1$ (in [3] it is assumed $a_2>a_1 \sqrt {2}$, but $a_2 > a_1 $
is sufficient).
Let us derive from (8), from equation (1) for  $v(x)$,
from the radiation condition for  $v(x)$, and from the estimate  $\|
v\|_{C^2(D_{12}^\prime)} \leq c$, the estimate:
$$
|v(x)| \leq c\epsilon^{d^{c^\prime}}, \quad x \in D_{12}^\prime,
\quad c_3 \rho \leq d \leq c_4 \rho, \quad c_3 > 0, \quad
d = \text{dist} (x, \Gamma_1^\prime), \tag 9
$$
If (9) is proved, then Theorem 1     follows.  Indeed, $|v(x)| = |v(s)
+ \nabla v \cdot (x - s)| = O(\rho) \leq c \epsilon^{\rho^{c^\prime}}$
if  $d$ satisfies (9).  Here we use: 
1) $v =u_2 - u_1 = u_2$  on  $\Gamma_1^\prime$, $|u_2| = O(\rho)$  on
$\Gamma_1^\prime$, since  $u_2 = 0$  on  $\tilde \Gamma_2$, and
$|\nabla u_2| \leq c$, 2) $|x - s| = O(\rho)$  if  $\text{dist} (x,
\Gamma_1^\prime) = O(\rho)$, and 3) $0 < c \leq |\nabla v| \leq \tilde c$
if  $d$ satisfies (9).  The last claim
follows from the continuity of  $\nabla v(x)$, smallness of  $\rho$, $\rho
(\delta) \to 0$  as  $\delta \to 0$, and the fact that  $|\nabla
u_j|_{\Gamma_j} \not= 0$  almost everywhere (otherwise, by the
uniqueness of the solution to the Cauchy problem for (1), one concludes
that  $u_j = 0$  in  $D_j^\prime$, which contradicts  (2), since, by
(2), $|u_j| \to 1$  as  $|x| \to \infty$).  Thus  $\ln \rho \leq
c\rho^{c^\prime} \ln \epsilon$, or $(\ast) \ \frac {\rho^{c^\prime}}{\ln
(\rho^{-1})} \leq c/\ln (\epsilon^{-1})$, where  $\rho$  and  $\epsilon$
are small numbers, $0 < \rho$, $\epsilon < 1$, $c,c^\prime > 0$,
and $c$ stands for different constants.  It
follows from  $(\ast)$  that  $\rho \leq \{ c/\ln
(\epsilon^{-1})\}^{\frac {1 + \omega}{c^\prime}}$, where  $\omega \to 0$
as  $\epsilon \to 0$.  From the definition (8) of  $\epsilon$, one gets
the estimate of Theorem 1.      Thus, the proof of Theorem 1     is
completed as soon as (9) is proved.
\enddemo
Our argument remains valid if $|v|=O(\rho^m)$ with some
$m,\, 0<m<\infty$.
Such an inequality is always true for a solution $v$ to elliptic
equation (1) unless $v\equiv 0$ (see [26, p.14]).
 
\demo{Proof of (9)} Since  $\| v\|_{C^{2,\mu} (D_{12}^\prime)} \leq c$,
$v(x)$ vanishes at infinity,
and  $v$  solves (1), one can represent  $v(x)$  in  $D_{12}^\prime$  by
the volume potential: $v(x) = \int_{D_{12}} g(x - y) f(y) dy$, $f \in
C^\mu (D_{12})$, $g(x) := \frac {\exp (i|x|)}{4\pi |x|}$.  The function
$|x - y| = [r^2 - 2r|y| \cos \theta + |y|^2]^{1/2} := R$  admits
analytic continuation on the complex plane  $z = r \exp (i\psi)$  to the
sector  $S_\phi : |\arg z| < \phi$, if  $z^2 - 2z|y| \cos \theta + |y|^2
\not= 0$  for  $z$  in this sector.  We use the branch of $R$ for which
$ImR \geq 0$, and  $ReR|_{Imz = 0} \geq 0$.  The argument of  $R^2 :=
z^2 - 2z|y| \cos \theta + |y|^2$  is defined so that it belongs to the
interval  $[0, 2\pi)$, so that the analytic continuation of  $g(x - y)$
to the sector  $S_\phi$  is {\it bounded} there.  It is crucial to have at
least boundedness of the norm  $(\dagger) \ \|
v\|_{C^1(D_{12}^\prime)}$.  Indeed, $(\dagger)$  implies that one can
extend  $v$  from  $D_{12}^\prime$  to  $D_{12}$  as  $C^1 ({\Bbb R}^3)$
functions.  This is true although the boundary  $\partial D_{12}$  may
be nonsmooth to the degree which prevents using the known extension
theorems (Stein's theorem, for example).  The way to go around this
difficulty is to extend  $u_1$  and  $u_2$  separately to  $D_1$  and
$D_2$  respectively, and then take  $v = u_2 - u_1$  as the extension.
If  $v \in C^1 ({\Bbb R}^3)$  satisfies the radiation condition and the
Helmholtz equation, and is  $C^2$  in the interior
and in the exterior of  $D_{12}$, then it is representable as a sum of
the volume and single-layer potentials, and our argument, which uses
analytic continuation, goes through.  Without this assumption the
argument is not valid and the conclusion fails, as the following example
shows.
 
\vskip .1in
 
\noindent {\bf Example 1:} Let  $D := \{ x : |x| \leq 1, x \in {\Bbb
R}^3\}$, $v = v_\ell := \frac {h_\ell^{(1)}(r)}{h_\ell^{(1)}(1)} Y_\ell
(x^0)$, where  $h_\ell^{(1)} (r)$  is the spherical Hankel function,
$Y_\ell (x^0)$  is the normalized in  $L^2 (S^2)$  spherical harmonic.
It is well known that  $h_\ell^{(1)} (r) \sim i \sqrt{\frac 1{(\ell +
\frac 12)r}} ( \frac {2\ell + 1}{er})^{\frac {2\ell + 1}2}$
as  $\ell \to \infty$  uniformly in  $1 \leq r \leq b$, $b < \infty$  is
arbitrary.  Therefore  $v_\ell \sim r^{-(\ell + 1)} Y_\ell
(x^0)$  as  $\ell \to \infty$.  In any annulus  ${\Cal A} := \{ x : 1 <
a_2 \leq r \leq b\}$, one has  $\| v_\ell\|_{L^2(A)} \leq c
a_2^{-(\ell + 1)} \to 0$  as  $\ell \to \infty$.  On the other
hand  $\| v_\ell \|_{L^2(S^2)} = 1$  for all  $\ell$.  Thus, for
sufficiently large  $\ell$  the solution  $v_\ell$  to Helmholtz
equation is as small as one wishes in the annulus  ${\Cal A}$, but it is
not small at the boundary  $\partial D$: for any  $\ell$  its  $L^2
(\partial D)$  norm is one.  The reason for the solution to fail to be
small on  $\partial D$  is that the  $C^1$  norm of  $v_\ell$  is
unbounded, as  $\ell \to \infty$, on  $\partial D$.

Let us continue the proof of (9). The function 
$v(r,x^0,\alpha)$, where $\alpha$ is the same as in (2),
 $x^0 := x/r$, and $r = |x|,$  admits an analytic
continuation to the sector  $S$  on the complex plane  $z$, $S := \{ z :
|\arg [z - r(x^0)]| < \phi \}$, $\phi > 0$, $r = r(x^0)$  is the
equation of the surface  $\Gamma_1$  in the spherical coordinates with
the origin at the point  $O$, and  $v(z,x^0,\alpha)$  is bounded in
$S$.  The angle  $\phi$  is chosen so that the cone  $K$  with the
vertex at  $r(x^0)$, axis along the normal to  $\Gamma_1^\prime$  at the
point  $r(x^0)$, and the opening angle  $2\phi$, belongs to
$D_{12}^\prime$.  Such a cone does exist because of the assumed
smoothness of  $\Gamma_j$.  The analytic continuation of this type was
used in [18].  It follows from (8) that  $\sup_{r \geq a_2} |v(r)| \leq
\epsilon$, and  $\sup_{z \in S} |v(z)| \leq c$, since  $Im [z^2 - 2z|y|
\cos \theta + |y|^2]^{1/2} \geq 0$ in $S$.  
From this and the classical theorem about
two constants [22, p.\ 296], one gets  $|v(z)| \leq c\epsilon^{h(z)}$,
where  $h(z) = h(z,L,Q)$  is the harmonic measure of the set  $\partial
S \setminus L$  with respect to the domain  $Q := S \setminus L$  at the
point  $z \in Q$.  Here  $L$  is the ray  $[a_2, +\infty)$, $\partial S$
is
the union of two rays, which form the boundary of the sector  $S$,
and of the ray $L$.  The
proof is completed as soon as we demonstrate that  $h(z) \sim
kd^{c^\prime}$  as
$z \to r(x^0)$  along the real axis, $d:=|z-r(x^0)|,$
 $k = \text{const} > 0$, $c =\text{const} > 0$.  
This, however, is clear: let $r(x^0)$ be the origin, and
denote $z-r(x^0)$ by $z$. If one maps
conformally the sector $S$  onto the half-plane
$Re z\geq 0$ using the map  $w =z^{c^\prime}$, $c^\prime = \frac \pi{2\phi}$, 
then the ray  $L$  is mapped
onto the ray  $L := [a_2^{c^\prime}, +\infty)$, and (see [22, p.\
293])  $h(z,L,Q) = h(z^{c^\prime}, L^\prime, Q^\prime)$,
where  $Q^\prime$  is the image of  $Q$  under the mapping  $z \mapsto
z^{c^\prime} = w$.  By the Hopf lemma [23, p.\ 34], $\frac {\partial
h(0, L^\prime, Q^\prime)}{\partial w} > 0$, $h(0, L^\prime, Q^\prime) =
0$,
so  $h(w, L^\prime, Q^\prime) \sim kw = kz^{c^\prime}$  as  $z \to 0$,
and (9) is proved.  Theorem 1 is proved.  \qed
\enddemo
 
\vskip .2in
 
\demo{Proof of Proposition 1} It is proved in [2, p.\ 183] that the set
$\{ u_N (s,\alpha)\}_{\forall \alpha \in S^2}$  is complete in  $L^2
(\Gamma)$.  This implies existence of a function  $\nu_\epsilon (\alpha,
\theta)$  such that
$$
\| \int_{S^2} u_N (s,\alpha) \nu_\epsilon (\alpha, \theta) d\alpha -
\frac {\partial \exp (i\theta \cdot s)}{\partial N_s}\|_{L^2(\Gamma)} <
\epsilon, \tag 10
$$
where  $\epsilon > 0$  is arbitrarily small fixed number, $N_s$  is the
exterior normal to  $\Gamma$  at the point  $s$, and  $\theta
\in M$  is an arbitrary fixed vector.  It is well known [1, p.\ 52],
that
$$
-4\pi A(\theta^\prime,\alpha) = \int_\Gamma \exp (-i\theta^\prime \cdot
s) u_N (s,\alpha)ds.  \tag 11
$$
Multiply (11) by  $\nu_\epsilon (\alpha,\theta)$, integrate over  $S^2$
and use (10), to get
$$
-4\pi \lim_{\epsilon \to 0} \int_{S^2} A(\theta^\prime, \alpha)
\nu_\epsilon (\alpha,\theta) d\alpha = \int_\Gamma \exp (-i\theta^\prime
\cdot s) \frac {\partial \exp (i\theta \cdot s)}{\partial N_s} ds.
\tag 12
$$
Note that
$$
\aligned
\int_\Gamma \exp (-i\theta^\prime \cdot s) \frac {\partial \exp (i\theta
\cdot s)}{\partial N_s} ds &= \frac 12 \int_\Gamma \frac {\partial \exp
[-i (\theta^\prime - \theta) \cdot s]}{\partial N_s} ds \\
&= \frac 12
\int_D \nabla^2 \exp (-i\lambda \cdot x)dx = -\frac {\lambda^2}2 \tilde
\chi_D (\lambda)
\endaligned \tag 13
$$
where the first equation is obtained with the help of Green's formula.
From (12) and (13) one obtains (5).  Proposition 1 is proved.  \qed
\enddemo
 
\vskip .2in
 
\demo{Remark 3} In [4]-[5] attempts are made to obtain stability results
for IOSP, but several errors invalidate the proofs in [4], [5] and
[6] related to stability for IOSP. Let us point out some of the errors.
Lemma 5, as stated in [4, p.\ 83], repeated as Lemma 4 in [5], claims
that if a solution to a homogeneous Helmholtz equation in the exterior
of a bounded domain  $D$  is small in the annulus  $R \leq |x| \leq R +
1$, $|v| \leq \epsilon$  in the annulus, then  $|v|_{\partial D} \leq
c|\log \epsilon|^{-c_1}$.  This is incorrect as Example 1 shows.  Lemma
3 in [4] is wrong (factor  $\rho^{2m}$  is forgotten in the argument).
In fact, stronger results have been published earlier [17],
[2], [3].  In [5] Lemma 2 is intended as a correction of Lemma 3
in [4] (without even mentioning [4]), but its proof is also wrong:
the factor $\rho^{2m}$ is not estimated.
  There are other mistakes in [5]
(e.g., the known asymptotics of Hankel functions in [5, p.\ 538] is
given incorrectly). In [6] these mistakes are repeated (p.\ 600).
 There are  claims in [6] that: a) there is a gap in the
 Schiffer's proof of the uniqueness theorem for
IOSP with the data  $A(\alpha^\prime, \alpha_0, k) \ \forall
\alpha^\prime \in S^2$, $\forall k > 0$, $\alpha_0 \in S^2$  is fixed
[6, p.\ 605], b) that Theorem 6 in [8] is incorrect, and the proof of
Lemma 5 in [8] contains a flaw [6, p.\ 588].  These claims are wrong,
and no justifications of the claims are given.  The remark concerning
Shiffer's proof in [6, p.\ 605, line 1] is irrelevant (see [1,pp.85-86]).
 It should be
noted that the arguments in [4]-[5] are based on the well known
estimates of Landis [9] for the stability of the solution to the Cauchy
problem, but no references to the work of Landis are given.
In [6] it is not mentioned that the concept of completeness of the set
of products of solutions to PDE (which is discussed in [6]) has been
introduced and widely used for the proof of the uniqueness theorems in
inverse problems in the works [2], [13], [19]-[21] (see also references
in [2], [13]).  In [24] and [25] two  theorems are announced
which contradict each other (Theorem 1 in [25] and Theorem 2 in [24]).
\enddemo
 
\subhead Acknowledgements \endsubhead
 The author thanks NSF for support and Prof. H.-D.Alber
for useful discussions.

\vfill
 
\noindent e-mail: ramm\@math.ksu.edu
 
\pagebreak

\enddocument